\documentclass[preprint,12pt]{elsarticle}

\usepackage{amssymb}
\usepackage{graphicx}
\usepackage{graphics}
\usepackage{rotating}
\usepackage{longtable}
\usepackage{lscape}
\usepackage{epsfig}
\usepackage{rotating}
\usepackage{ifthen}
\usepackage{multicol}
\usepackage{amssymb}
\usepackage{xspace}

\newcommand{\Otwo}{\ensuremath{\rm O_2}\xspace}
\newcommand{\Ntwo}{\ensuremath{\rm N_2}\xspace}
\newcommand{\Argon}{argon\xspace}
\newcommand{\Ar}{\ensuremath{\rm Ar}\xspace}
\newcommand{\meth}{methane\xspace}
\newcommand{\He}{\ensuremath{\rm He}\xspace}
\newcommand{\water}{\ensuremath{\rm H_2O}\xspace}
\newcommand{\AMU}{\ensuremath{\rm u}\xspace}

\newcommand{\ppm}{\ensuremath{\rm \cdot10^{-6}~g/g}\xspace}
\newcommand{\oneppm}{\ensuremath{\rm 10^{-6}~g/g}\xspace}
\newcommand{\ppb}{\ensuremath{\rm \cdot10^{-9}~g/g}\xspace}
\newcommand{\oneppb}{\ensuremath{\rm 10^{-9}~g/g}\xspace}
\newcommand{\ppt}{\ensuremath{\rm \cdot10^{-12}~g/g}\xspace}

\newcommand{\Xe}{\ensuremath{\rm xenon}\xspace}

\newcommand{\degree}{\ensuremath{^{\circ}}\xspace}

\newcommand{\figwidth}{\columnwidth}
\newcommand{\ife}[3]{\ifthenelse{\equal{#1}{#2}}{#3}{{}}}

\begin{document}

\begin{frontmatter}


\title{A simple high-sensitivity technique for purity analysis of xenon gas}

\author{D.~S.~Leonard\fnref{DL}}
\author{A.~Dobi}
\author{C.~Hall}
\author{L.~Kaufman}
\author{T.~Langford}
\author{S.~Slutsky}
\author{Y.~R.~Yen}

\address{Department of Physics, University of Maryland, College Park MD, 20742 USA}

\fntext[DL]{Now at University of Seoul, Seoul, South Korea}

\begin{abstract}

We report on the development and performance of a high-sensitivity purity-analysis technique for gaseous xenon. The gas is sampled at macroscopic pressure from the system of interest using a UHV leak valve. The xenon present in the sample is removed with a liquid-nitrogen cold trap, and the remaining impurities are observed with a standard vacuum mass-spectroscopy device. Using calibrated samples of xenon gas spiked with known levels of impurities, we find that the minimum detectable levels of  \Ntwo, \Otwo, and methane are 1\ppb, 160\ppt , and 60\ppt respectively. This represents an improvement of about a factor of 10\,000 compared to measurements performed without a cold trap.

%Xenon in both liquid and gas forms has recently become a medium of choice for radiation and particle detection in a range of experiments.  Trace impurities in the xenon impact the performance of detectors in multiple ways.  Electronegative impurities, especially oxygen, reduce the electron life-time and thus achievable drift distance within liquid xenon.  Radioactive impurities can produce unwanted background signals or may be purposely added for calibration.  For these reasons it is often critical for experiments to have instrumentation to monitor the levels and types of impurities within the xenon. We report here on the development and performance of a high sensitivity xenon gas analyzer consisting of a standard residual gas analyzer enhanced by a liquid nitrogen cold trap.
\end{abstract}

\end{frontmatter}

\section{Introduction}
\label{intro}
 
Liquid xenon has become a common detection medium for many particle, astrophysics, and even medical imaging studies~\cite{AprileXeRev09}.  Large gaseous xenon detection systems are also in development~\cite{NEXT}.  A typical xenon detection system detects deposited energy by collecting some combination of ionization charge and scintillation light. Charge collection requires electron lifetimes in the xenon sufficiently long so that a large fraction of ionization electrons can drift to the detector anode.   Electronegative impurities at the levels of about \oneppb reduce the electron attenuation length to several tens of centimeters depending on the electric field in the liquid xenon~\cite{Bakale}.  Higher levels thus produce relevant limits to detector sizes or prevent charge collection all together. Impurities are also responsible for the attenuation of vacuum ultraviolet light and therefore present a problem for xenon scintillation detectors~\cite{Baldini}. In other situations, trace levels of impurities may be desirable.  For example, methane isomers such as $\rm ^{14}CH_4$ and $\rm CH_3T$ could be used as internal beta calibration sources for large liquid-xenon experiments.  For these reasons, it is often critical that these experiments have instrumentation to allow for monitoring of the xenon purity.  

Existing xenon purity monitors typically measure the electron lifetime in the liquid phase via charge attenuation measurements~\cite{puritymonitors,argonpurity}. For ionization detectors, the electron lifetime is exactly the quantity of concern, and a precise measurement of its value allows for corrections to be made to the charge collection signal. Impurities may be present in the commercial xenon source in relatively high concentrations, but such impurities can be easily removed by a gas purifier while filling the detector. Of more concern are impurity sources which act downstream of the purifier, such as outgassing from detector components and plumbing, or small leaks in the system which have escaped detection. One strategy for overcoming these problems is liquid phase purification~\cite{Barabash85,Bolotnikov}, whereby the purifier and the detector are in effect unified into a single device. Although liquid phase purification technology is promising, it has not yet been widely adopted in xenon detectors. Instead, the most common method employed at present is gas phase purification, typically with a heated zirconium getter, coupled with continuous re-circulation of the xenon through the detector volume. This arrangement is capable of achieving xenon purities which are adequate for present day experiments and has become a standard technique. In practice, however, the electron lifetime is sometimes observed to be small, despite the best efforts of the experimenters. In this case, the electron lifetime measurement alone does not provide much information regarding the source of the problem, be it excessive outgassing, leaks, or a failure of the purifier.

It would therefore be useful to have a complementary purity monitoring technique which can operate in the gas phase downstream of the purifier, and which can identify the chemical species of the impurities which are responsible for the poor electron lifetime. One possible method is atmospheric-pressure ionization mass spectroscopy (API-MS), which is capable of detecting impurities at concentrations of less than \oneppb. Unfortunately, this technology is out of reach for the average experimenter due to cost. On the other hand, labs which host ultra-high vacuum (UHV) systems, including most xenon detector labs, are usually equipped with a residual gas analyzer (RGA), a mass spectroscopy device which operates at pressures of $10^{-5}$~Torr or less. Xenon gas from the detector may be introduced into the RGA through a capillary tube or leak valve, and the chemical composition of the gas, including any impurity species, can be ascertained. However, this technique by itself has limited usefulness because the dynamic range of the RGA is only six orders of magnitude at best. Under these circumstance, impurities can only be detected at concentrations greater than one part per million, far above the part-per-billion concentrations which are relevant for ionization detection. In essence, the RGA is saturated by the overwhelming xenon pressure, and this limits the sensitivity of the device to impurities.

We report here on the development of a method to improve the sensitivity of the RGA to impurities by limiting the partial pressure of the xenon with a liquid nitrogen cold trap. The method relies on the fact that the physical properties of many common impurity species differ sufficiently from that of xenon such that they can be separated with a cold surface. This separation allows the gas sampling device (leak valve or capillary tube) to operate at vastly higher rates without risk of saturating the RGA.

In this paper, we present the results of our tests with methane, nitrogen, and oxygen, three impurity species which have freezing points well below that of xenon (161 K). (We expect that impurities with freezing points above xenon, such as carbon dioxide and water, could also be detected with a more sophisticated trap-and-release method, but we have not attempted such detection at this time.) At liquid nitrogen temperature (77~K), the vapor pressure of solid xenon is 1.8~mTorr.  \Ntwo and \Otwo both have vapor pressures greater than or equal to 1~atm at 77~K.  In the case of methane, the freezing point is 91~K, but its vapor pressure at 77~K is 8~Torr~\cite{MethVapor}, still far above the xenon vapor pressure.  It is thus reasonable to expect that these gases would pass through a liquid nitrogen trap at much higher partial pressures than the xenon itself. In fact, we find that the cold-trap/RGA method is capable of detecting methane, for example, at concentrations as low as 60 parts-per-trillion. This represents an improvement of more than four orders of magnitude over a standalone RGA method.

\section{The cold trap hardware}
\label{analyzer}

\begin{figure}[htb]\centering
\includegraphics[width=\figwidth]{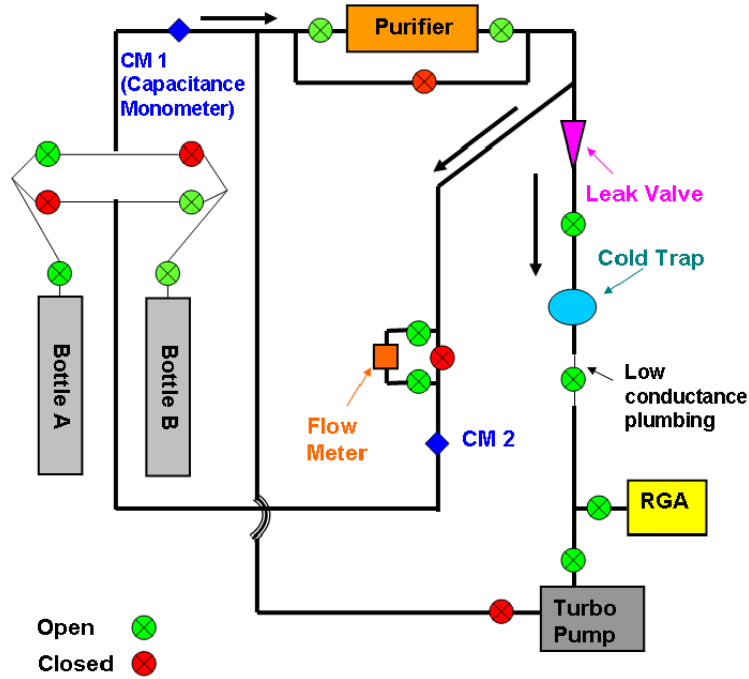}
\caption{Xenon handling system: There is leak valve to a cold trap at the output of the purifier. The cold trap is followed by an RGA and a turbo pump. Xenon can be cryopumped from one storage bottle to the other to flow gas through the primary xenon system as desired. As drawn, the valves are configured to supply a constant pressure of gas (regulators are not shown) from bottle A as gas is sampled by the cold-trap/RGA analysis system.}
\label{fig:system}
\end{figure}

A hardware schematic of the analysis system is shown in Fig.~\ref{fig:system}. We sample gas from our multi-purpose gas handling system with a controllable leak valve (Kurt J. Lesker, LVM series, Part Number VZLVM940R) which reduces the pressure from one atmosphere to a fraction of a torr and allows control of the flow rate into the cold trap.  After passing through the leak valve the gas travels through about 1~m of 0.95~cm ID SS piping where it then expands into a horizontal SS pipe of 3.81~cm diameter.  The 3.81~cm pipe extends for about 33~cm before making a 90\degree bend downward into a liquid-nitrogen Dewar vessel (24~cm ID).   The pipe extends downward into the dewar about 50~cm, making a 180\degree\xspace ``U''-shaped bend of about a 6~cm radius. The pipe returns up and out of the dewar.  The output of the cold trap then connects through about 13~cm of 0.95~cm ID SS pipe containing a shut-off valve (open during normal operation).  The exit of the pipe connects to about 1.5~m of 3.81~cm SS plumbing before splitting to an SRS-200 RGA and a 70~L/s turbo pump.  A cold cathode gauge monitors the gas pressure between the RGA and the turbo pump. The low conductance of the 0.95~cm cold-trap exit pipe further reduces the output gas pressure to $\rm 8\cdot10^{-6}~Torr$, just under the maximum operating pressure of the RGA.  Because the output conductance should ideally be reproducible, the shut-off valve is probably not well suited for conductance control.   For our trap the output conductance was set as needed by the plumbing design alone and the valve was fully open during operation.  

\section{Operation}
\label{sec:ops}

For typical operation of the xenon gas purity analyzer we use a storage bottle regulator to maintain the primary xenon handling system at a pressure of about 1050~Torr.  While pumping on the cold trap and monitoring masses of interest with the RGA, we fill the cold-trap dewar with nitrogen to about 25~cm above the bottom of the trap bend.  We have found that gas analysis results are insensitive to the liquid nitrogen level.  After a brief wait for the trap to cool down, we then inject xenon gas into the trap by opening the leak valve.  The LVM series leak valve has a peculiar marking system to label valve opening positions. Fractional rotations are marked in increments from 0 to 50, with 50 being a full rotation of the valve control knob.  We refer here to decimal factions of real knob turns, not the markings printed on the knob itself.   For initial gas injection we typically open the valve 1.00 turns.   We have estimated the flow rate of Xenon gas through the leak valve by monitoring pressure loss in a closed system.  Table~\ref{tab:valve} lists gas flow rates at various input leak valve positions.  We have found valve hysteresis in the form of 10\% to 20\% variations in flow rate depending on which direction the valve was last turned.  In order to reduce this error, we always set the flow by opening the valve to the desired position without turning past it.

\begin{table}
\begin{tabular}{cc}
Valve\newline Position &  Leak Rate [$\rm Torr\cdot L/min$] \tabularnewline
\hline
1.0  & $.05^*$ \tabularnewline
1.1  & $.24^*$ \tabularnewline
1.15 & $.52^*$ \tabularnewline
1.2  & 1.5  $\pm$ 0.1 \tabularnewline
1.25 & 2.9  $\pm$ 0.5 \tabularnewline
1.3  & 7.1  $\pm$ 0.6 \tabularnewline
1.35 & 17.4 $\pm$ 1.5 \tabularnewline
1.4  & 43   $\pm$ 3 \tabularnewline
\hline
\end{tabular}
\caption{Rate of xenon gas flow through the cold trap input leak valve for different valve positions.  The measurements were obtained with an input pressure of about 1050~Torr. Uncertainties listed are related to pressure measurements for a single test of the valve; they do not include errors associated with the reproducibility of the valve opening.\newline
$\rm ^*$ Rates for valve openings below 1.2 turns were inferred by assuming that the helium partial pressure response is proportional to the xenon flow rate below 1.2 turns.}
\label{tab:valve}
\end{table}

We use the 132 \AMU RGA channel as a monitor for the xenon partial pressure, representing about a quarter of the various xenon isotopes.  We typically monitor all pressures using the channel electron multiplier (CEM) of the RGA for increased sensitivity.  After opening the leak valve to 1.00 turns the xenon RGA pressure immediately rises sharply from a typical background level of about $5\cdot10^{-12}$~Torr, leveling off within a couple of minutes to a reproducible pressure of about $5\cdot10^{-7}$~Torr.   

After establishing the xenon vapor pressure, the leak valve can be opened further or shut off entirely without impacting the xenon RGA pressure, supporting the claim that the xenon output pressure, is, under normal conditions, controlled by the solid xenon ice in the trap and not by the gas flow rate through the leak valve.   This characteristic is, of course, the key to maximizing analysis sensitivity.  Further increases of the input leak rates result in increased RGA signals of other impurities  while the xenon pressure remains constant.  In our studies these impurities include \Otwo, \Ntwo, methane, argon, and helium.   Limitations and backgrounds are discussed in more detail in Secs.~\ref{sec:limitations} and~\ref{sec:backgrounds}. 

To measure impurities at or below approximately \oneppm, we open the leak valve to 1.4 turns.  After any change in the leak valve position we typically see that partial pressures reported by the RGA require about three or four minutes to stabilize.  At this point the RGA partial pressure is simply read off as a relative measure of the concentration of the impurity in the \Xe.  

The xenon pressure typically falls by about 10\% during the first 30 to 60 minutes of operation after initially opening the leak valve to 1.00 turns.  We find evidence that gas analysis results may be more reproducible, at the level of a few percent, after allowing some time for the xenon pressure to stabilize.  Furthermore we find fluctuations in tracer-normalized results from day to day, after gas handling operations, on the level of about 5\% of the signals.

After finishing an analysis session, the solid xenon in the cold trap is recovered by closing the output valve of the cold trap, opening the input valve, and removing the Dewar vessel to allow the trap to warm up while cryopumping into one of the xenon gas storage bottles.

\subsection{Tracer gases}
\label{sec:tracer}

Errors due to  non-reproducibility of input flow rates, discussed above, can be reduced to unobservable levels by normalizing all data to partial pressure readings of \Argon, which was present in our stock \Xe , or \He which we added at a concentration of about 8\ppb to use as a tracer.  This process also compensates for small changes in the pressure at the input to the leak valve, which can cause non-linear RGA responses, and provides a good general diagnosis tool to monitor the health of the apparatus. Furthermore, normalization to a tracer serves to monitor intentional changes made to the leak valve.  We find that the partial pressures of all species are, to a good approximation, proportional to the input flow rate.  It follows that normalization of signals of any species of interest to that of a tracer gas with a constant concentration compensates the analysis for large intentional changes in the input flow rate used to control gas consumption or sensitivity.   

When testing purification systems it can be beneficial to change the input flow rate to adjust analysis sensitivity as impurity concentrations change.  Particularly it can be useful to improve sensitivities by briefly increasing the leak rate to levels which, if  sustained for long times, would cause the cold trap to clog with, or develop unsafe levels of, xenon ice.  Noble gas tracers are particularly suited to purification studies because they are generally unaffected by purifiers. This technique is discussed more in Ref.~\cite{purifier}.

\begin{figure}[]\centering
\includegraphics[width=\figwidth]{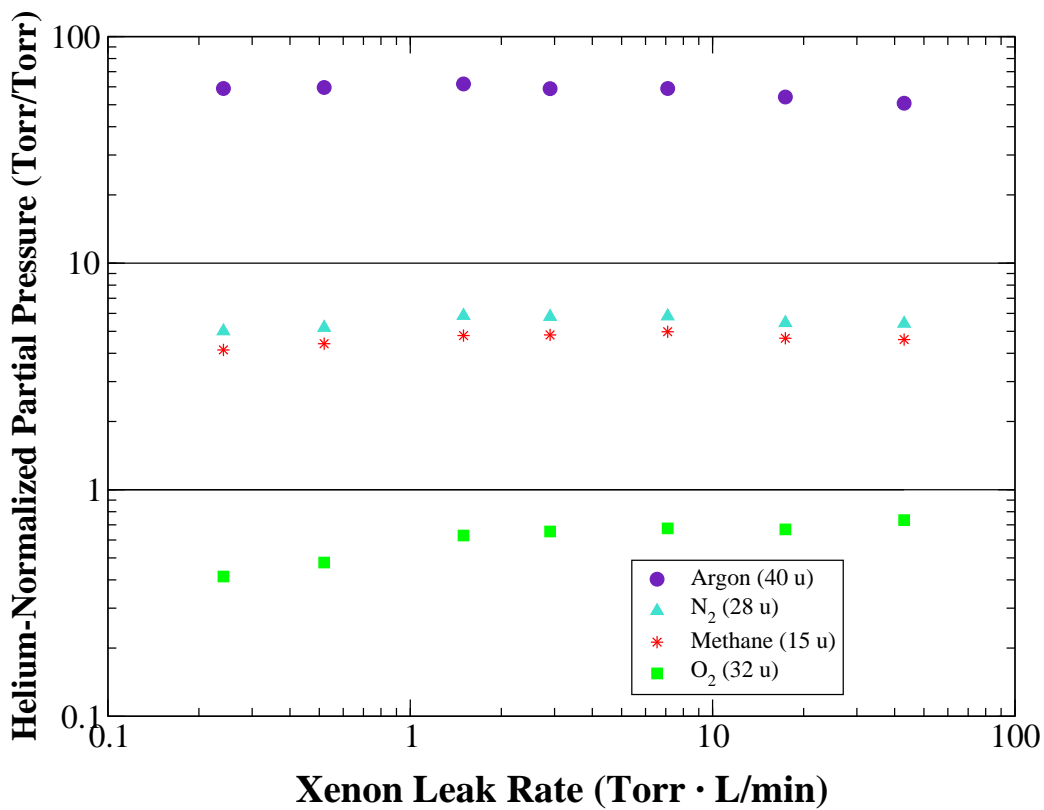}
\caption{Ratio of RGA partial pressures of various xenon contaminants to that of \He for a range of gas leak rates through the cold-trap input leak valve.   The leak rate was inferred from the valve position along with valve calibration data in Table~\ref{tab:valve}.  For leak rates below $\rm 1~Torr\cdot L/min$ the rate was extrapolated using the helium partial pressure. The values and scale differences between species are specific to our particular gas mixture.  Ideally the helium normalized partial pressures would be independent of the leak rate.  Variations of as much as a factor of 1.8 are present in these data, but above $\rm 1 Torr\cdot L/min$, where we operate our cold trap, fluctuations are only about 20\% of the values. These data can be used to derive species dependent correction factors for our system.}
%Uncertainties in the leak rate translate to to 20\% uncertainties in the normalized values, correlated between all species, as the data seem to indicate. It is visibly evident that fluctuations can be reduced by normalizing results instead to the partial pressure of a tracer gas, particularly He.
\label{fig:valve}
\end{figure}

Before using \He as a tracer gas we must first confirm that the partial pressures of the impurities scale with input flow rate in the same way that He does.  Figure~\ref{fig:valve} shows the ratios of partial pressures of various impurities to the partial pressure of \He at multiple flow rates.  The ratios are found to have total fluctuations of about 10\% to 20\% at flow rates above $\rm 1~Torr\cdot L/min$, with \Ar being the worst, and both \meth and \Ntwo having total fluctuations below 10\%. This variation is a few times worse, with a systematic trend, if data at lower flow rates are included.  The trend at low flow rates can be corrected in data analysis, but is probably still an acceptable error level for many applications.  Also, we see no obvious reason, even with significantly smaller traps, to operate at such low flow rates. All further data reported here were collected at a valve position of 1.4 turns; therefore no correction is required.  Variations above a leak rate of $\rm 1~Torr\cdot L/min$ become about three times worse if data are instead normalized to the leak rate inferred from the valve position.

We notice that argon partial pressures do not track the other species particularly well, especially at high flow rates, possibly because, for the levels of argon in our xenon, its signal is near saturation in the RGA. For this reason \He is a preferred tracer gas in these conditions. 

%Normalization of RGA signals to a tracer gas partial pressure is optional, but can have other advantages. Changes in the pressure at the input side of the leak valve will result in changes in the observed RGA signals.  Changes in the tracer gas partial pressure will indicate this condition, and, for small pressure changes, can be used to correct for it. However, we have found that partial pressures in the RGA do not change linearly as a function of changing xenon input pressure.    The best use of the tracer may simply be to provide an extra known diagnostic signal to verify the healthy operation of the apparatus.  For example, on one or two occasions we found that the RGA response had changed from one day to the next, and that sensitivities of various impurities relative to \Xe had changed by as much as factors of two.  Although these changes were obvious, the presence of the extra tracer can help in diagnosing such occurrences.

Use of the tracer gas is optional and can have disadvantages.  We have found that freezing and thawing the xenon can, not surprisingly, result in very significant separation of the tracer gases from the xenon.   Even after warming the xenon to its gas phase, these gases can remain separated for long times.  This seems especially true if the xenon gas is allowed to boil off into long plumbing sections rather than bottles.  To use the tracers for relative normalization during a measurement sequence, one must insure that the tracer gases have been uniformly mixed throughout all of the xenon being tested.  We found that we could achieve constant tracer signals by allowing our storage bottles to sit overnight after they were warmed.

\subsection{Limitations}
\label{sec:limitations}

The ultimate performance and sensitivity of the cold-trap/RGA apparatus is limited by background levels in the plumbing and RGA (See Sec.~\ref{sec:backgrounds}), by interferences from impurities of neighboring masses (See Sec.~\ref{sec:Interference}), by the presence of uninteresting impurities (such as argon or helium) which may saturate the RGA, and by practical constraints on achievable and sustainable input flow rates.  The latter two of these limitations are discussed in this section. 

The input flow rate through the leak valve to the trap may, of course, be limited by plumbing or the available gas feed rate from the primary xenon system.  Moreover, it can also be limited by the capacity of the cold trap to safely contain solid xenon coupled with the need to have the trap operate for a practical length of time.   With a previous prototype cold trap made from 0.95~cm ID SS plumbing, we found that we could typically operate the trap for about 1.5 hours at a flow rate of $\rm 2.9~Torr \cdot L/min$  before a solid xenon ice plug would completely block the flow path in the trap. This condition is detectable by dramatic drops in all partial pressures seen in the RGA.  Furthermore, to gain additional sensitivity, it is desirable to open the leak valve further, which results in a proportionate decrease in the amount of measurement time available before the trap is blocked. With the present trap design, using the 3.81~cm ID plumbing, it is possible to operate the trap for several hours with the leak valve opened 1.4 turns.  We have not found the limit to the operating time in this mode.  However, it is very important to note that at this valve position gas is flowing into the trap at the rate of about $\rm 2\,700~Torr\cdot L/hour$.  With a trap volume of roughly 1.5~L, significant and potentially hazardous pressures could develop if the trap is operated for long periods of time, or at higher input rates, and is then warmed up without providing sufficient pressure relief and gas recovery mechanisms.  

As already noted, the dynamic range of the RGA limits sensitivity by establishing a maximum measurable ratio between partial pressures of xenon and impurities.  It is also true though that the same limit applies to the partial pressures of any two impurities so that an overpressure of any impurity can limit the sensitivity to all other impurities.  We found that a few times \oneppm of \Argon present in our \Xe gas resulted in the \Argon and \Xe partial pressures becoming approximately equal for flow rates of about $\rm 45~Torr\cdot L/min$.  Since the trap was operated such that xenon always produced the maximum acceptable pressure in the RGA, then above this flow rate, \Argon would add significantly to the total pressure and saturate the RGA.  Thus the input rate and signal gain relative to \Xe could not be increased by opening the leak valve further.  For a cleaner gas not having this limitation, it may be possible to achieve better overall sensitivities.  Higher flow rates are probably not easily sustainable for extended times but could feasibly be used for brief periods in order to sample with higher sensitivity.  For our set-up we expect that we could achieve, without modifications, about ten times higher flow rates, and thus ten times improved sensitivities.

\section{Calibration data}
\label{sec:data}

In order to calibrate the partial pressure measurement in terms of absolute impurity concentration, and to understand the ultimate sensitivity of the apparatus, we spiked our xenon gas supply with known levels of impurities. We first purified the xenon gas twice through a SAES (Monotorr PS4-MT3) heated zirconium getter at a flow rate of 5~standard-liters per minute (SLPM). Studies of the purifier performance are reported in Ref~\cite{purifier}. A sample of the impurity species of interest was prepared by filling a small section of evacuated plumbing with the impurity gas. The volume and pressure of this section of plumbing were then measured to determine the mass of the gas sample. To achieve very small quantities, the sample was allowed to expand into a larger section of evacuated plumbing of known volume.  A small volume was again separated from the rest of the larger volume by closing a valve.
The large volume was then evacuated to remove the excess gas. Once the sample was prepared, it was swept into the xenon supply bottle, using xenon as a carrier gas, and was then left overnight. 

 \Ntwo and \Otwo were, in some cases, added to the \Xe simultaneously, but \meth was studied first, separately, and after a dedicated purification to reduce interferences discussed in Sec.~\ref{sec:Interference}. The mixtures were stored in one of our two storage bottles shown in Fig.~\ref{fig:system}.  For \meth we collected data at concentrations ranging from 250\ppt to \oneppm.  Measurement procedures are described in Sec.~\ref{sec:ops}.  For \meth analysis we monitored the RGA signal at 15 \AMU since 16 and 14 have higher backgrounds and interferences (See Secs.~\ref{sec:backgrounds} and~\ref{sec:Interference}).

\begin{figure}[htb]\centering
\includegraphics[width=\figwidth]{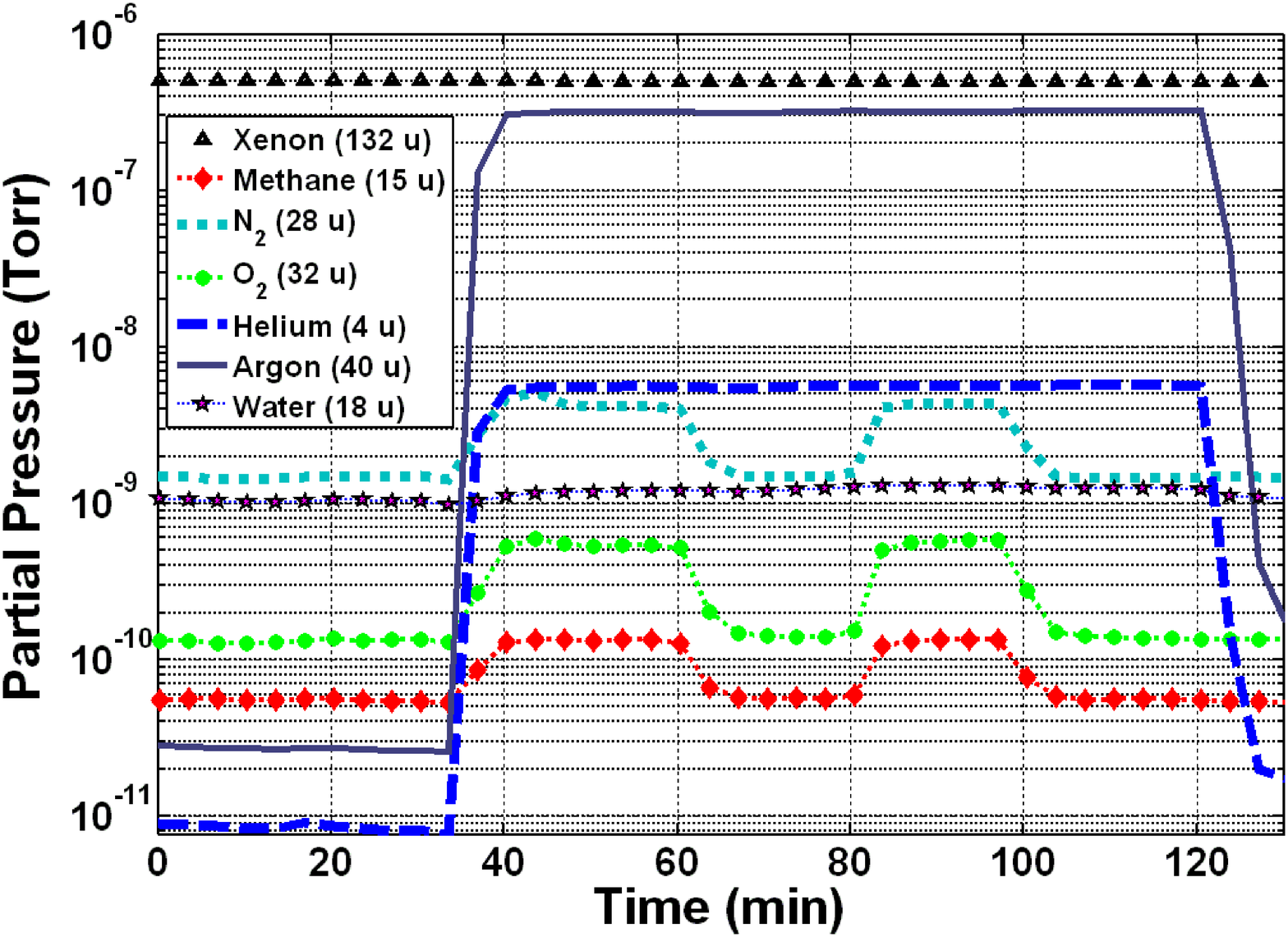}
\caption{RGA partial pressure responses for measurement of impurities in xenon prepared with 250\ppt of methane.  Helium was added at about 8\ppb as a tracer gas.  Argon was present in the xenon from the supplier.  \Ntwo and \Otwo were introduced after purification, at levels of 18\ppb and 5\ppb respectively, by leaks or outgassing in our xenon system. At 0 minutes xenon ice and thus a xenon vapor pressure had already been establish in the cold trap, but the leak valve had then been closed. The leak valve to the cold trap was re-opened at about 35 minutes, sampling gas after it flowed through the purifier bypass plumbing.  The output xenon vapor pressure is unaffected by the input flow of xenon due to the action of the cold trap. At 60 minutes the gas was rerouted through the purifier before reaching the leak valve.  This cycle was repeated.  The effect of purification on the methane as well as on \Ntwo and \Otwo is clear.  Interference from \Ntwo and \Otwo contribute to about 15\% of the observable methane change. Helium and Argon were unaffected by the purifier and effectively monitored minor fluctuations in the gas flow rate through the leak valve.}
\label{fig:250ppt}
\end{figure}

For the lowest \meth concentration we verified the signal by first analyzing the purified gas before injecting the methane.  We then analyzed the methane mixture while flowing gas from one bottle while cryopumping into the other at a flow rate of 1~SLPM.  The gas was sent through the purifier bypass while sampling the 250\ppt mixture at the downstream leak valve.  We then closed the bypass and sent the gas through the purifier, making adjustments to maintain constant pressure at the leak valve. When alternating between bypass and purify mode, we observed the \meth signal rise and fall as expected (see Fig.~\ref{fig:250ppt}), showing a clear indication that the observed signal was related to a removable contaminant in the gas.  Interference from \Otwo and \Ntwo contributed to about 15\% of the methane response.  See Secs.~\ref{sec:backgrounds} and~\ref{sec:Interference} for more details about backgrounds and interferences.

\begin{figure}[htb]\centering
\includegraphics[width=\figwidth]{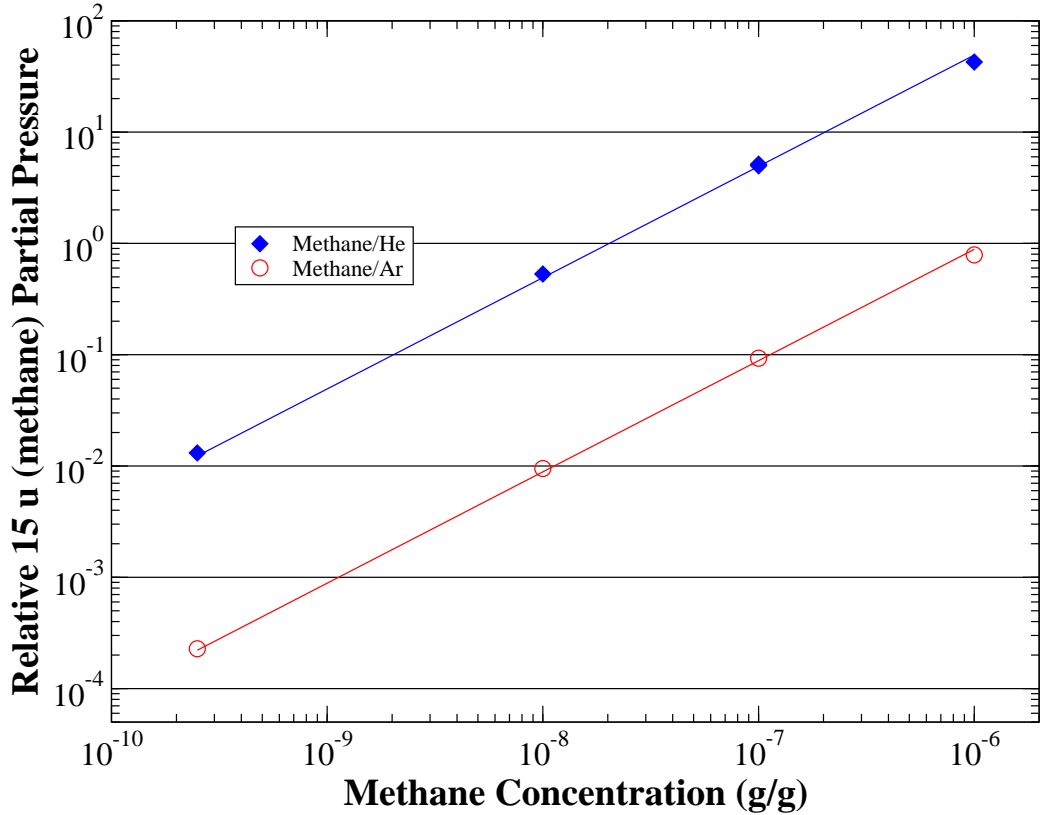}
\caption{Normalized methane partial-pressures in the RGA as a function of methane concentration in the xenon gas.  The mass 15 component of the methane signal is shown normalized to the partial pressures of both \He and \Ar, each having fixed and arbitrary concentrations in our xenon throughout these measurements.  Linear fits passing through the origin are shown for visual reference.  Backgrounds and interferences have been subtracted. Uncertainties are not shown. The values are averaged over a couple of minutes of RGA sampling, producing statistical errors of well less than 1\%. The point at the lowest concentration, 250\ppt, has an additional 15\% uncertainty from subtraction of interference signals and backgrounds (See Secs.~\ref{sec:Interference} and~\ref{sec:backgrounds}.  We find about 5\% total variation in all results from day to day tests.} 
\label{fig:calibration}
\end{figure}

Fig.~\ref{fig:calibration} shows results for all \meth calibration measurements with backgrounds and interferences subtracted and with the data normalized to  each of the fixed \He and \Ar partial pressures (See Sec.~\ref{sec:tracer} for tracer-normalization details). The data at 1\ppm and 100\ppb represent an average of data taken multiple times on each of two or three days.  Results of repeated measurements agree within a total spread of about 5\%.  All data shown were taken with the leak valve open 1.4 turns, corresponding to a flow rate of $\rm43~Torr\cdot L/min$.  We can calculate a calibration constant from each point by taking the ratio of the normalized signal to the concentration.  For the helium-normalized data, we find that for all but the highest ( 1\ppm) calibration point, the standard deviation in the calibration constants is 2.4\% with a total spread of 7\%. The 1\ppm calibration however is 20\% lower than the mean of all of the calibrations at lower concentrations.  Data taken for \oneppm of methane sampled at a lower flow rate, with the leak valve opened to 1.2 turns, does fall within the spread of the other calibrations.  We believe that the discrepancy at 1.4 turns probably comes from the RGA becoming nearly saturated by the addition of the extra methane partial pressure. This is consistent with the similar observation noted in Sec.~\ref{sec:tracer} for responses of argon (which also had concentrations on the order of \oneppm) at high flow rates. 

Calibration responses for \Otwo concentrations ranging from 600\ppt to \oneppm are shown in Fig.~\ref{fig:Ofig:O2calibration}. The \He tracer gas was not present during this calibration and the data were thus normalized only to the response from the fixed \Ar partial pressure.  Collection of the lowest \Otwo calibration point was simplified, in comparison to methane, due to the lack of any interference signal at 32 u.  As was the case for methane, the calibration constant for \Otwo was found to be fairly stable, with a standard deviation of about 8\% over the entire calibration range.  The \Otwo calibration data was performed several weeks after the methane calibration. At a time between these calibrations the RGA response suddenly changed a little for no known reason.  We found that such changes, which we have observed on two other occasions, had no effect on the linearity or sensitivity of the analysis system.

Presently we have calibrated the \Ntwo response of our apparatus at only one concentration, specifically \oneppm.  To extrapolate from this single point calibrations, we rely on the assumption that the RGA response for \Ntwo is approximately proportional to the concentration. Given that both \Otwo and methane scale linearly with concentration we expect the \Ntwo signal to respond linearly as well.  By comparing data at 1.2 and 1.4 leak-valve turns, we have determined that the RGA saturation effect noticed for methane is not present for \Ntwo and \Otwo. We attribute this to the fact that, at the same mass concentration,  \Ntwo and \Otwo have lower partial pressures than methane.     Table~\ref{tab:1ppm} compares the results for \Ntwo, \Otwo and \meth at \oneppm.  Detection sensitivities are discussed in Sec.~\ref{sec:backgrounds}.

\begin{figure}[htb]\centering
\includegraphics[width=\figwidth]{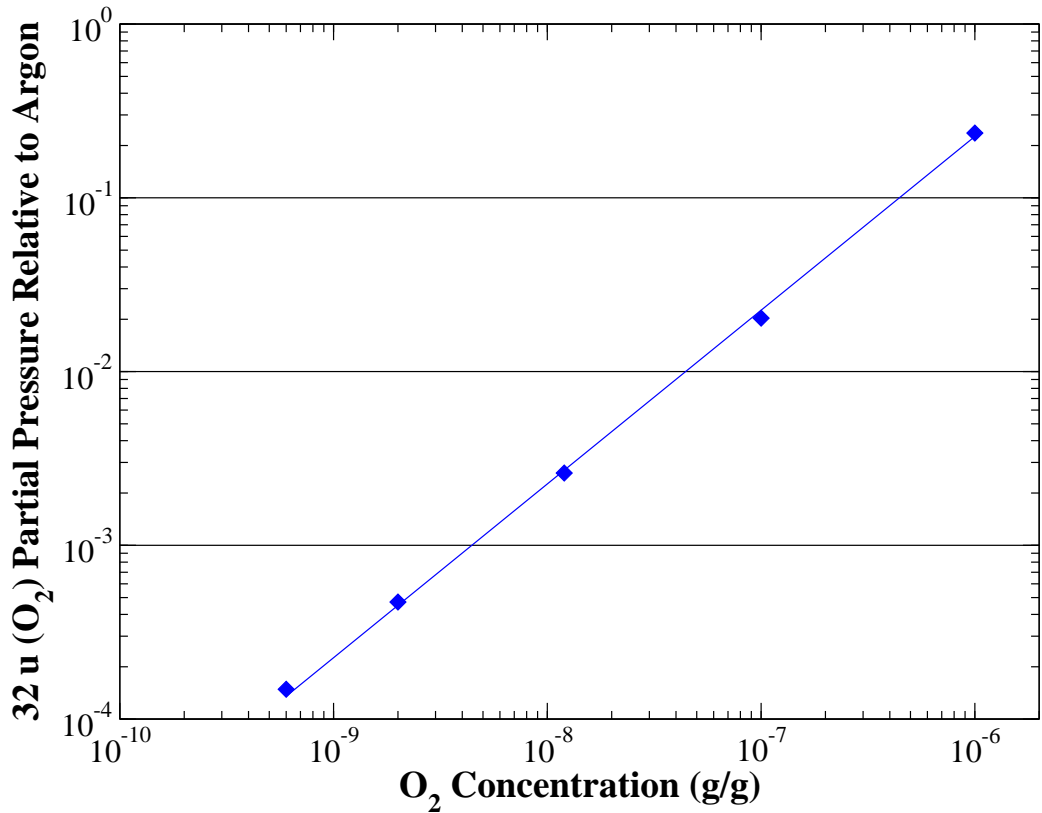}
\caption{Normalized \Otwo partial-pressures in the RGA as a function of \Otwo concentration in the xenon gas.  The mass 32 RGA signal is shown normalized to the partial pressures of \Ar, having fixed and arbitrary concentration in our xenon throughout these measurements.  A linear fit passing through the origin is shown for visual reference. Backgrounds have been subtracted.  Uncertainties are not shown. The values are averaged over a couple of minutes of RGA sampling, producing statistical errors of well less than 1\%.} 
\label{fig:O2calibration}
\end{figure}

\begin{table}[h]
\begin{tabular}{lccc}
~&Partial Pressure (Torr)&\tabularnewline
\hline
\hline
Methane (15 u) & $2.5\cdot10^{-7}$ \tabularnewline
\Ntwo (28 u)   & $1.5\cdot10^{-7}$ \tabularnewline
\Otwo (32 u)   & $9.3\cdot10^{-8}$ \tabularnewline
\end{tabular}
\caption{Comparison of observed RGA signals for \oneppm concentrations of \meth, \Ntwo, and \Otwo represented by partial pressures of mass 15 \AMU, 28 \AMU and 32 \AMU respectively for a xenon flow rate into the cold trap of $\rm43~Torr\cdot L/min$.  Statistical uncertainties are below 1\% of the values.  Systematic errors in the partial pressures are dominated by input flow rate uncertainties of as much as 20\% (See Sec.~\ref{sec:ops}).  Normalization of these partial-pressures to that of a  tracer  gas, as discussed in Sec.~\ref{sec:tracer}, can reduce systematic uncertainties.}
\label{tab:1ppm}
\end{table}

\section{Backgrounds and sensitivities}
\label{sec:backgrounds}
Backgrounds and interference signals typically result in sensitivities which are worse than those expected from the quoted  RGA dynamic range alone. We consider any RGA signal which is independent of the input gas mixture to be a background. These signals are generated by out-gassing from the cold trap and the vacuum plumbing.  

%Backgrounds
%Interferences are dependent on the RGA mass resolution. 
We find, in agreement with the RGA documentation, that the RGA mass peaks fall by about 90\% in a deviation of 1~\AMU. This is important for understanding backgrounds, since signal contributions may come from the mass of interest or from neighboring masses. However, for quantitative analysis of backgrounds, all contributions are stable after sufficient pumping time;  thus regardless of the source, the background signal of the mass of interest must simply be measured and subtracted.  In practice, some error can arise because of the peak finding algorithm as discussed later in this section.

Without interferences between RGA signals from the different contaminants in the input gas, the sensitivity of our analysis system is limited to about 10\% of the background levels. It is best to measure background partial pressures with the cold trap immersed in liquid nitrogen and with a xenon vapor pressure established. We find that background levels can rise slightly when any gas pressure is present in the trap. After establishing the xenon vapor pressure at a valve position of 1.0 turns, we measure the background levels  by closing the input leak valve.  Purification tests have verified that the background levels obtained this way are consistent with those observed by analyzing gas from the output of the purifier.  This provides support for two conclusions: The purifier is effective even at very low concentrations (See Ref~\cite{purifier} for more details), and that closing the input leak valve is a valid way to measure signal backgrounds. There is one caveat.  We find that if high concentrations of \meth are analyzed, then the subsequent \meth background level in the cold trap increases. If no further \meth is injected, the background level returns to its baseline value over a time scale of about one hour.  

\subsection{\Otwo and \Ntwo backgrounds}

Background levels change on long time scales depending on the history and cleanliness of the trap and vacuum plumbing.  Table~\ref{tab:backgrounds} lists background partial pressures for various impurities for the day of the 250\ppt methane calibration. Equivalent concentrations are also shown by using the calibration data in Sec.~\ref{sec:data} to interpret the partial pressures.  For  \Ntwo a single calibration point was used along with an assumption that the partial pressure is directly proportional to the concentration, an assumption which is found to be approximately valid for \Otwo and methane calibration data. With statistical uncertainties of only a couple of percent, impurities in \Xe resulting in deviations of 10\% from background levels are clearly detectable as positive observations, corresponding to ultimate sensitivities of $160\cdot10^{-12}$ and $970\cdot10^{-12}$ for \Otwo and \Ntwo respectively. Results could almost certainly be improved by reducing unneeded plumbing such as long corrugated vacuum hoses, by adding higher pump-out conductance to the trap, and by baking. 

\subsection{Methane backgrounds}

We found that the detection of small methane signals can be problematic in some situations. Methane is a difficult case because the signal at 15 u can be hidden by the tail of the 16 u peak due to atomic oxygen. This situation is made more difficult because the RGA software does not perform a true peak search at 15 u. Instead, it simply assumes that the highest reading between 14.7 u and 15.3 u is due to the 15 u signal. However, when methane levels are very low, the highest value will occur at 15.3 u due to the neighboring peak at 16 u. Therefore, before a methane signal will be reported, it must exceed the atomic oxygen reading at 15.3 u. As as result, we find that the smallest detectable methane signal is 30\% of the methane background, assuming the typical oxygen background level. We apply appropriate corrections and uncertainties to results at or near the detection threshold. A more sophisticated analysis of the RGA raw data could improve upon this method, but we have not attempted to implement such a technique at this time.

\begin{table}[h]
\begin{tabular}{cccc}
            & Background Partial & Equivalent          & Limit of \tabularnewline
RGA channel & Pressure [Torr]    & Concentration [g/g] & Detection [g/g]\tabularnewline
\hline
\hline
4  u (helium)  & $ 8.30\cdot10^{-12} $ & & \tabularnewline
40 u (argon) & $2.80\cdot10^{-11}$ &  & \tabularnewline
15 u (methane) & $5.6\cdot10^{-11}$ & $190\cdot10^{-12}$ & $57\cdot10^{-12}$ \tabularnewline
28 u (\Ntwo) & $1.5\cdot10^{-9}$  & $9.7\cdot10^{-9}$ & $970\cdot10^{-12}$ \tabularnewline
32 u (\Otwo) & $1.4\cdot10^{-10}$ & $1.6\cdot10^{-9}$ & $160\cdot10^{-12}$ \tabularnewline
\end{tabular}
\caption{Backgrounds for each species and the resulting limits of detection. Results can vary from day to day.  Measurements were made when the input gas was purified or the input was valved off. Backgrounds are also shown in units of equivalent concentration, where the value shown is valid for a leak rate of 43 Torr$\cdot$L/min.  For \Otwo and \Ntwo, the limit of detection is 10\% of the background level. For methane, the limit of detection is 30\% of the background level. See text for details.}
\label{tab:backgrounds}
\end{table}

\section{Interference}
\label{sec:Interference}

The background levels listed in Sec.~\ref{sec:backgrounds} provide a measure of the best purity limits measurable by our analysis system, at least for the plumbing cleanliness on the day the measurements were made.  However, positive detections above these levels, although certainly indicative of an impurity, must be interpreted carefully since, as described above in Sec.~\ref{sec:backgrounds}, signals from different impurities may not be resolvable.  In particular we found that \Ntwo and \Otwo concentrations can interfere with \meth detection.  The primary RGA signal for \meth is at a mass of 16 \AMU.  We chose to monitor the nearly equal signal at 15 \AMU because 16 \AMU also corresponds to elemental oxygen produced in the RGA from \Otwo and \water.  Even so, the mass 15 signal is contaminated by the tail of the mass 16 peak at about 10\% of the mass 16 peak signal level.  A similar interference is produced from the mass 14 signal arising from the presence of \Ntwo.  As previously stated, the 15 \AMU background from stable out-gassing of \Ntwo and \Otwo and \water is measurable and subtractable.  However if \Ntwo and \Otwo exist in the xenon, they result in 15 \AMU signals above background levels which will disappear when the gas is purified or when the input valve is closed, thus mimicking a real \meth contamination. Furthermore, due to the peak finding algorithm of the RGA, and the potential lack of a peak at 15 \AMU, the stability and reproducibility of the interference signal should be questioned.

Indeed our mixture of Xe and \meth did contain small levels of \Otwo and \Ntwo from an unknown source. Therefore our calibration measurements at 250\ppt required particular care. \Otwo presented the primary source of problematic interference. We purified the gas multiple times and stored it in a bottle overnight before performing measurements.  Nevertheless, we observed that \Otwo and \Ntwo levels in the stored gas increased over several days, although we were unable to locate any leak in our system with a helium leak-check procedure.
To minimize interference from this effect, we performed all measurements within 30 hours of the end of purification.  Furthermore, we tested, on multiple occasions, the 15 \AMU interference signal after purifying the \Xe gas multiple times, storing it overnight, and without adding \meth.  Analysis was done in a method identical to the 250\ppt \meth calibration by flowing stored gas from the bottle and alternately passing the gas directly to the leak valve input or first through the purifier.  We did see a small change in the 15 \AMU signal at the level of about $\rm 1.5\cdot10^{-12}~Torr$, corresponding to a false \meth reading of 35\ppt.  We found that for data taken within one or two days of purification this result was reproducible within a factor of about two.  To be conservative we chose our lowest \meth calibration point to be several times higher than this level.  We subtracted off the result of the interference measurement taken just a few hours earlier while verifying stability of \Ntwo and \Otwo during those few hours. 

For gas void of \Ntwo, and \Otwo, positive measurements made at lower levels can be trusted.  Even for gas containing our observed levels of impurities, observations of signals at or near the level of the interference can still be confidently used to set upper limits at levels below 100\ppt, and for gas directly out of our purifier, we can set methane limits at $\rm < 60\ppt$ or slightly better.

\section{Summary}

We have constructed and calibrated a simple and inexpensive device to analyze xenon gas impurities with sensitivities below \oneppb for \Ntwo, \Otwo, and \meth.  For xenon gas taken directly from the output of a SAES zirconium purifier we can set limits on \meth concentrations of $\rm <60\ppt$.  We have investigated the limitations and backgrounds of our implementation of this technique and have proposed simple methods and conditions which could yield results with even better sensitivities.  The technique uses only relatively inexpensive hardware, most of which is already in use at most xenon gas experimental facilities.  We believe that this analysis method is a novel tool that can greatly help to understand propagation and removal of impurities in xenon based detection experiments and devices.

\bibliographystyle{elsarticle-num}
\bibliography{purifier}

\end{document}